\documentclass[12pt,preprint]{aastex}
\usepackage{amssymb}
\usepackage{graphicx}

\def\ds{\displaystyle}

\def\rin{r_{\rm in}}
\def\rout{r_{\rm out}}

\def\lea{\mathrel{<\kern-1.0em\lower0.9ex\hbox{$\sim$}}}
\def\gea{\mathrel{>\kern-1.0em\lower0.9ex\hbox{$\sim$}}}

\shorttitle{Spectroscopic Study of NGC 6822 ESCs} \shortauthors{Hwang et al.}

\begin{document}

\title{Spectroscopic Study of Extended Star Clusters\\ in Dwarf Galaxy NGC
6822\altaffilmark{*}}

\author{Narae Hwang\altaffilmark{1,2,6}, Hong Soo Park\altaffilmark{3},
Myung Gyoon Lee\altaffilmark{3}, Sungsoon Lim\altaffilmark{3}, 
Paul W. Hodge\altaffilmark{4}, Sang Chul Kim\altaffilmark{1}, Bryan Miller\altaffilmark{5}, and Daniel Weisz\altaffilmark{4}}
\affil{$^1$ Korea Astronomy and Space Science Institute \\
776 Daedeokdae-Ro Yuseong-Gu, Daejeon 305-348, Korea}
\affil{$^2$ National Astronomical Observatory of Japan \\2-21-1 Osawa
Mitaka, Tokyo 181-8588, Japan} 
\affil{$^3$ Astronomy Program, Department of Physics and Astronomy,\\
Seoul National University, Seoul 151-747, Korea} 
\affil{$^4$ Department of Astronomy, University of Washington, Seattle, WA 98195-1580, USA} 
\affil{$^5$ Gemini Observatory, Casilla 603, La Serena, Chile} 

\altaffiltext{*}{Based on observations obtained at the Gemini Observatory, which is operated by
the Association of Universities for
    Research in Astronomy, Inc., under a cooperative agreement with the
    NSF on behalf of the Gemini partnership: the National Science
    Foundation (United States), the Science and Technology Facilities
    Council (United Kingdom), the National Research Council (Canada),
    CONICYT (Chile), the Australian Research Council (Australia),
    Minist\'{e}rio da Ci\^{e}ncia, Tecnologia e Inova\c{c}\~{a}o (Brazil)
    and Ministerio de Ciencia, Tecnolog\'{i}a e Innovaci\'{o}n Productiva
    (Argentina)}
\altaffiltext{6}{e-mail: nhwang@kasi.re.kr}

\begin{abstract}
We present a spectroscopic study of the four extended star clusters (ESCs) in  NGC 6822 based on the data obtained with Gemini Multi-Object Spectrograph (GMOS) on the Gemini-South 8.1m telescope. 
Their radial velocities derived from the spectra range from $-61.2 \pm 20.4$ km s$^{-1}$ (for C1) to $-115.34 \pm 57.9$ km s$^{-1}$ (for C4) and, unlike the intermediate age carbon stars, they do not display any sign of systematic rotation around NGC 6822. The ages and metallicities derived using the Lick indices show that the ESCs are old ($\geq 8$ Gyr) and metal poor ([Fe/H]$\lesssim -1.5$). NGC 6822 is found to have both metal poor ([Fe/H]$\approx -2.0$) and metal rich ([Fe/H]$\approx -0.9$) star clusters within 15 arcmin (2 kpc) from the center, while only metal poor clusters are observed in the outer halo with $r \geq 20$ arcmin (2.6 kpc).
The kinematics, old ages, and low metallicities of ESCs suggest that ESCs may have accreted into the halo of NGC 6822. 
Based on the velocity distribution of ESCs, we have determined the total mass and the mass-to-light ratio of NGC 6822: $M_{N6822} = 7.5^{+4.5}_{-0.1} \times 10^{9} M_{\odot}$ and $(M/L)_{N6822} = 75^{+45}_{-1} (M/L)_{\odot}$. It shows that NGC 6822 is one of the most dark matter dominated dwarf galaxies in the Local Group.
\end{abstract}


\section{Introduction}

Globular clusters (GCs) have played an important role as tracers of stellar populations under the context that GC is a simple system of stars that can be represented by an assumption of simple stellar population. 
Recently, it has been found that GCs have much more diverse sub-populations than conventionally believed: GCs with multiple stellar populations, faint fuzzy clusters, and extended star clusters (ESCs), etc. This leads to an issue how these new populations of GCs could challenge the long-standing usage of GCs as tracers of stellar populations and kinematics in galaxies. On the other hand, for ESCs that are characterized by systematically larger size than typical GCs, this kind of challenge could be another opportunity for a new tracer if ESCs turn out to be associated with some peculiar properties of galaxies that may allow a peek into their evolutionary history. One example is an argument that ESCs might be associated with dwarf galaxy accretion events (e.g., \citealt{mac04,hwa11,hwa12}). However, this requires thorough studies on the physical, chemical, and kinematic properties of ESC population.

One simple complication to such studies is that ESCs are usually faint and are discovered in rather distant galaxies, sometimes more than several Mpc away from the Milky Way (MW). This is a serious challenge because spectroscopic observations are required for the reliable determination of physical properties, which is feasible only with nearby star clusters. In this regard, the ESCs in NGC 6822, one of the nearest dwarf galaxies in the Local Group, are ideal targets for the spectroscopic studies. There are at least five ESCs in NGC 6822 discovered to date \citep{hwa05,hwa11,hux13}.

For a small and isolated dwarf irregular galaxy, NGC 6822 is known to have a relatively large stellar halo populated with intermediate age carbon stars with $1 \sim 3$ Gyr \citep{let02} as well as red giant stars with $\geq 1$ Gyr \citep{lee05}. Combined with bright and young stars concentrated on the central body as well as HI disk-like structure \citep{komi03,deb00}, NGC 6822 has complex stellar structure that extends out to about $30$ arcmin, well beyond the extent of a central optical body. This has led to a suggestion that NGC 6822 might have undergone merger-like interactions rather recently, which have significantly increased the star formation rate within the last 3 Gyr \citep{gal96c}.

To understand physical conditions and extent of primordial NGC 6822, we need to concentrate on old stellar components that still pervade throughout the galaxy. The spheroidal or halo of a galaxy is an ideal for this purpose since it is mainly composed of old stars with much less contamination by bright and young stars usually found near the galaxy center. A recent study by \citet{sib12} shows that a significant population of intermediate age AGB stars is found out to 28 arcmin, corresponding to 4 kpc, from the center of NGC 6822 in projected distance, which is consistent with the results of \cite{let02} and \cite{lee05}. However, there is no clear evidence on the boundary of NGC 6822 halo and not much information on the properties of stellar components in the very outer part of its halo.

The ESCs discovered in NGC 6822 have expanded the boundary of its stellar halo even further out to 77 arcmin or 10.5 kpc from the center of galaxy.  
Since ESCs are star clusters with many observational advantages over individual stars, this provides an excellent opportunity to investigate the physical and kinematic properties of outer stellar components as well as to explore the extent of the galaxy. From the photometric study based on the CFHT MegaCam image data, we have shown in \citet{hwa11} that those new ESCs are probably old and metal poor populations. However, the results suffer uncertainties caused by insufficient spatial resolution and shallow photometric depth of the ground-based image data as well as the age-metallicity degeneracy inherent to the photometric CMD method.

In this study, we present the first result of the integrated spectroscopic study for four ESCs among those five discovered in
NGC 6822 based on the observations made with GMOS at Gemini-South. We assume the same distance to
each ESC as NGC 6822 based on the result of CMD analysis \citep{hwa11} and adopt $(m-M)_0 = 23.35 \pm 0.02$, an average of distance moduli measured by using RR Ryrae \citep{cle03}, Cepheid \citep{pie04}, and NIR TRGB \citep{cio05}. This translates to about 470 kpc in physical distance to NGC 6822 where $1\arcmin$ corresponds to about 0.132 kpc in projected scale.

\section{Observations and Data Reduction}
\label{datred}

The target ESCs in NGC 6822 were discovered in our previous imaging survey of NGC 6822 with CFHT MegaCam and the detailed information on these clusters including the coordinates, magnitude, and size is listed in \citet{hwa11}.
The spectroscopic observations of the four ESCs in NGC 6822 were made using the Gemini Multi-Object
Spectrograph (GMOS) of Gemini-South 8.1m telescope in queue mode operation under the program ID
GS-2011B-Q-16 (PI: N. Hwang) between August and October 2011.
We used the GMOS in long-slit mode with B600\_G5323 grating in which each target is placed at the
central 0.75 arcsec wide slit in each exposure. 
The exposures were taken in multiples of 700 sec (for C1) or 1800 sec (for C2, C3, and C4) and the total
integrated exposure times run from 2800 sec for C1 to 3 hours for C2/C4 and 4 hours for C3.
We dithered each exposure by $0.5-1.0$ arcsec to prevent a few bright stars from dominating the whole spectra, which is necessary since the target ESCs are partially resolved into their member stars. Each exposure was binned by 2 pixels in the spectral direction and by 4 pixels in the spatial direction to improve the signal of the obtained spectra. The wavelength coverage of the spectra is about 3900--6500 {\AA} with a spectral resolution of about 0.92 {\AA} per pixel, safely including major absorption features of interest such as Ca H+K, H$\beta$, Mgb, and some iron lines including Fe5270 and Fe5335.

We also obtained spectra of two old GCs in NGC 6822, Hubble VII (H VII) and Hubble VIII (H VIII)
for the comparison with target ESCs, and four MW GCs (NGC 6624, M12, M15, M107) for the calibration of Lick indices using the same observational settings. The integrated exposure times for these clusters range from 1 min to 1 hour depending on the surface brightness. For the flux calibration, we observed LTT7379 as our spectral standard. Each science exposure was preceded and followed by flat field and CuAr arc exposures with the same observational settings.

\begin{figure}
 \plotone{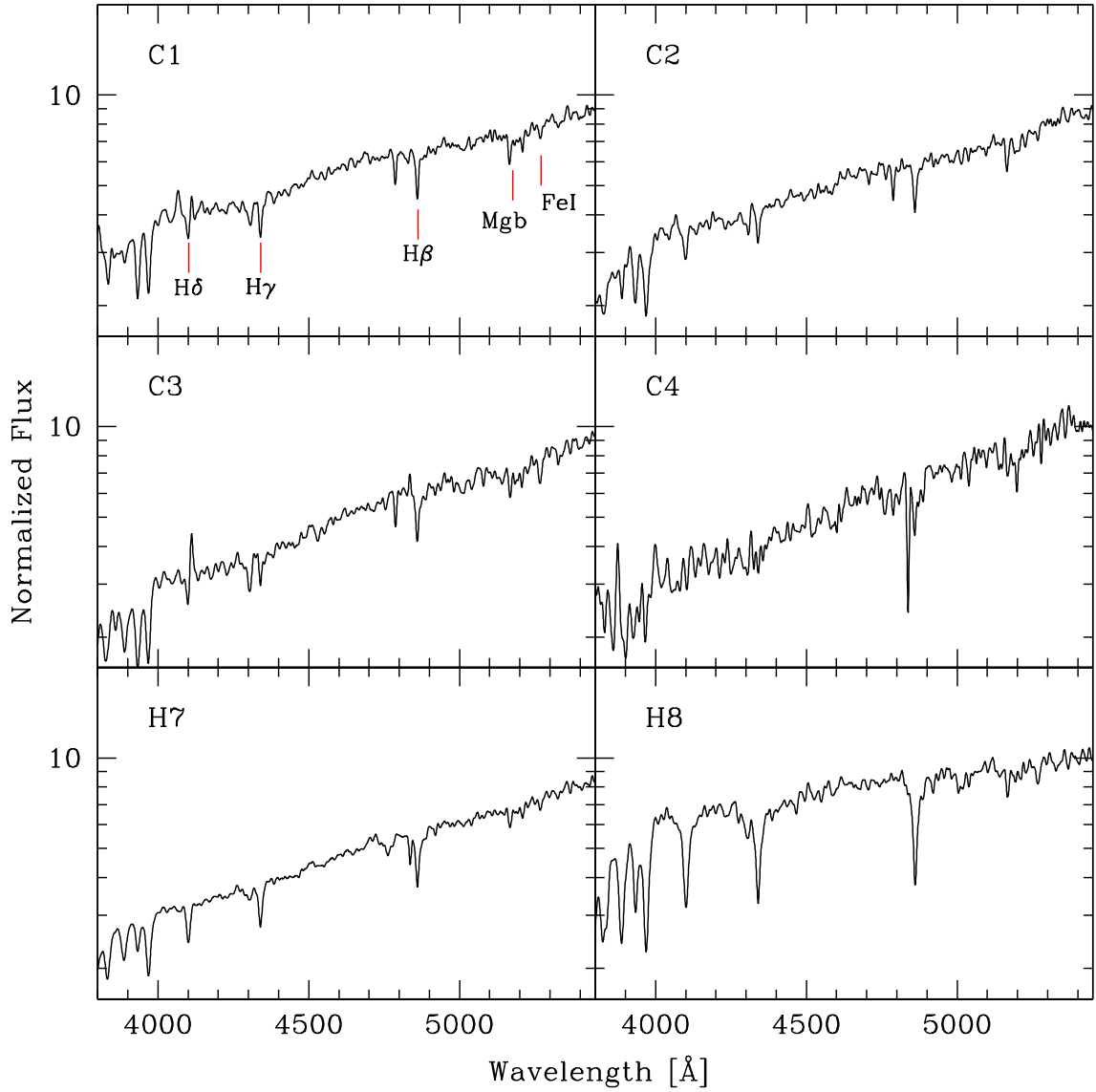}
 \caption{The spectra of NGC 6822 ESCs (C1-C4) as well as Hubble VII (H7) and Hubble VIII (H8). The spectra are smoothed with the Lick resolution and some major Lick index lines are marked for C1.}
\label{spectra}
\end{figure}

The reduction of data was carried out with the Gemini package in IRAF \footnote{IRAF is distributed by the National Optical Astronomy Observatory, which is operated by the Association of Universities for Research in Astronomy (AURA) under cooperative agreement with the National Science Foundation.}.
First, we used $gprepare$ to prepare the raw data for further reduction and $gsreduce$ to
flatfield, to correct overscan, to trim, and to mosaic the science images. The wavelength
calibration was established by using $gswavelength$ task on CuAr arc taken together with
science data, and then, the derived wavelength solution was applied to the corresponding science
data by using $gstransform$. The calibrated science spectra were combined to make a single
data and then, the combined data were traced and extracted by using $gsextract$ to make one
dimensional spectral data. 
The finally reduced spectra with several major absorption features are shown Figure \ref{spectra}.

\section{Radial Velocity Measurements}
\label{rv}

We have derived the radial velocities of target clusters by using $fxcor$ task in IRAF $rv$ package. The spectra of four MW GCs (NGC 6624, M12, M107, and M15) were used as templates for the correlation with target spectra over the wavelength range of $4800-5500$ \AA~ where several prominent absorption lines including H$\beta$ and Mgb are available. The radial velocities of MW GCs were taken from the catalog by \citet{har96} (2010 version) and are in the range of -107.0 km/s for M15 to 53.9 km/s for NGC 6624. Each target spectrum was correlated with template spectra of four MW GCs and then the derived velocities were weight-averaged with errors to calculate the final radial velocity.

The determined radial velocities range from $-61.2 \pm 20.4$ km/s for C1 to $-115.3 \pm 57.9$ km/s for C4 and are listed in Table 1.
For H VII and H VIII, the radial velocities are determined to be $-64.5 \pm 20.6$ km/s and $-46.9 \pm 31.2$ km/s, respectively. These velocities of H VII and H VIII are in good agreement with the radial velocity $-57.0 \pm 2.0$ km/s of NGC 6822 listed in the NASA/IPAC Extragalactic Database (NED) within the measurement errors.

\section{Lick Index Measurements}
\label{lick1}

The standardized Lick index system \citep{wor94,wor97,tra98} is a set of the absorption line strength measures for low-resolution spectra that can be compared to theoretical stellar population models to derive the ages, metallicities, and [$\alpha$/Fe] of target system. There are 25 Lick indices, running from CN$_1$ to TiO$_2$, of which measurement wavelength range was defined by \citet{wor94}. 

We measured the Lick indices in the spectra of four ESCs and two Hubble clusters in NGC 6822 as well as four MW GCs using GONZO\footnote{GONZO is available upon request from T.H. Puzia.}, a code for Lick index measurement described in \citet{puz02}. 
GONZO measures the line index of each Lick feature according to the definition by \citet{wor94}. We used the sky spectrum for each target with GONZO to simulate the uncertainty of the Lick index measurements through 100 Monte Carlo realizations. For each realization, GONZO measures the indices again in the simulated spectra, and the 1 standard deviation of the measured index is adopted as the index uncertainty.

For the calibration of the measured Lick index values, we compared the index measurements of four MW GCs, such as NGC 6624, M12, M107, and M15 with the standard Lick index values listed in \citet{tra98} and \citet{kun02}. Then, the zero point offsets between standard Lick and the measured indices were calculated in the following way: $Index(Lick) = Index(GMOS) + offset$. We determined the zero point offsets for 24 indices that were measured in more than two MW GC calibrator spectra. The offsets and their rms are listed in Table \ref{tab2}. The finally derived 24 Lick indices and their errors for four ESCs and two Hubble clusters in NGC 6822 are listesd in Tables \ref{tab3a} and \ref{tab3b}, respectively.

\section{Age and Metallicity Determination}
\label{agemetal}

For the reliable determination of age and metallicity from the Lick indices, we need high S/N for major indices such as H$\beta$, Mg2, Mgb, Fe5270, and Fe5335, where S/N value is measured at the continuum part of the corresponding Lick indices.
The S/N for those major Lick indices is between 15 and 60 for C1--C3, H7, and H8. However, for C4, the S/N values for the same indices are about 5--14, which leads to relatively large errors. 

We have used two methods for the age and metallicity determination of star clusters using Lick indices: a model grid method and a $\chi^2$ minimization method.

\subsection{Lick index model grid method}

We have used the Lick index model grid method (the grid method hereafter) to determine the age and metallicity of target clusters. This method compares the measured and calibrated Lick indices with the index grids calculated from the theoretical SSP models. In this study, we use the grids spanned by [MgFe]$^\prime$ versus Balmer lines such as H$\beta$, H$\delta$, and H$\gamma$ lines to estimate the metallicity and age \citep{tho04}. 
The composite index [MgFe]$^\prime$, defined as [MgFe]$^\prime$ $= \sqrt{Mgb \times (0.72 \times Fe5270 + 0.28 \times Fe5335)}$, is a good tracer for metallicity with little sensitivity to [$\alpha$/Fe], while Balmer lines are a good tracer of age \citep{tho03}. 
For the age and metallicity for a star cluster, we adopted a weighted average of values derived from the model grids of [MgFe]$^\prime$ versus H$\beta$, H$\delta$, and H$\gamma$ lines, respectively \citep{puz04}. 

However, each theoretical model grid still depends on [$\alpha$/Fe] that can be, in principle, derived from the grid of Mg2 versus $<Fe>$, where $<Fe> = (Fe5270+Fe5335)/2$, which is also a function of age. We have tried to determine [$\alpha$/Fe] using the grid of Mg2 versus $<Fe>$ based on the technique used by \citet{par12} that iterates between these model grids to finally determine the age, metallicity, and [$\alpha$/Fe] for target clusters. After some exhaustive tests with this approach, we have found that the derived [$\alpha$/Fe] values include rather large uncertainties that do not allow further physical application. Therefore, we decided to use the age and metallicity values determined by applying the theoretical model with [$\alpha$/Fe]=0.3, the average value for MW GCs, to the grid of [MgFe]$^\prime$ and Balmer lines.

\subsection{$\chi^2$ minimization method for Lick indices}

Instead of using a few major Lick indices, we have also applied a `$\chi^2$ minimization method' ($\chi^2$ fit method hereafter) that utilizes all available line indices to fit the model index lines to determine the age and metallicity from the input spectra. This method was suggested by \citet{pro04} as a more robust method for determining physical parameters of globular clusters.
We applied this method to our target star clusters using the same SSP models with [$\alpha$/Fe]=0.3 as used in the grid method. 
We also tested the fit result by varying [$\alpha$/Fe] values, but could not find any sign of significant improvement.
The total number of lines used for each cluster is about 15 with best signal available for the corresponding target. The error of each parameter was estimated by calculating the 1$\sigma$ significance level above the minimum point in the $\chi^2$ distribution. 

\section{Results}

\subsection{Kinematics of NGC 6822 ESCs}
\label{kin0}

\begin{figure}
 \plotone{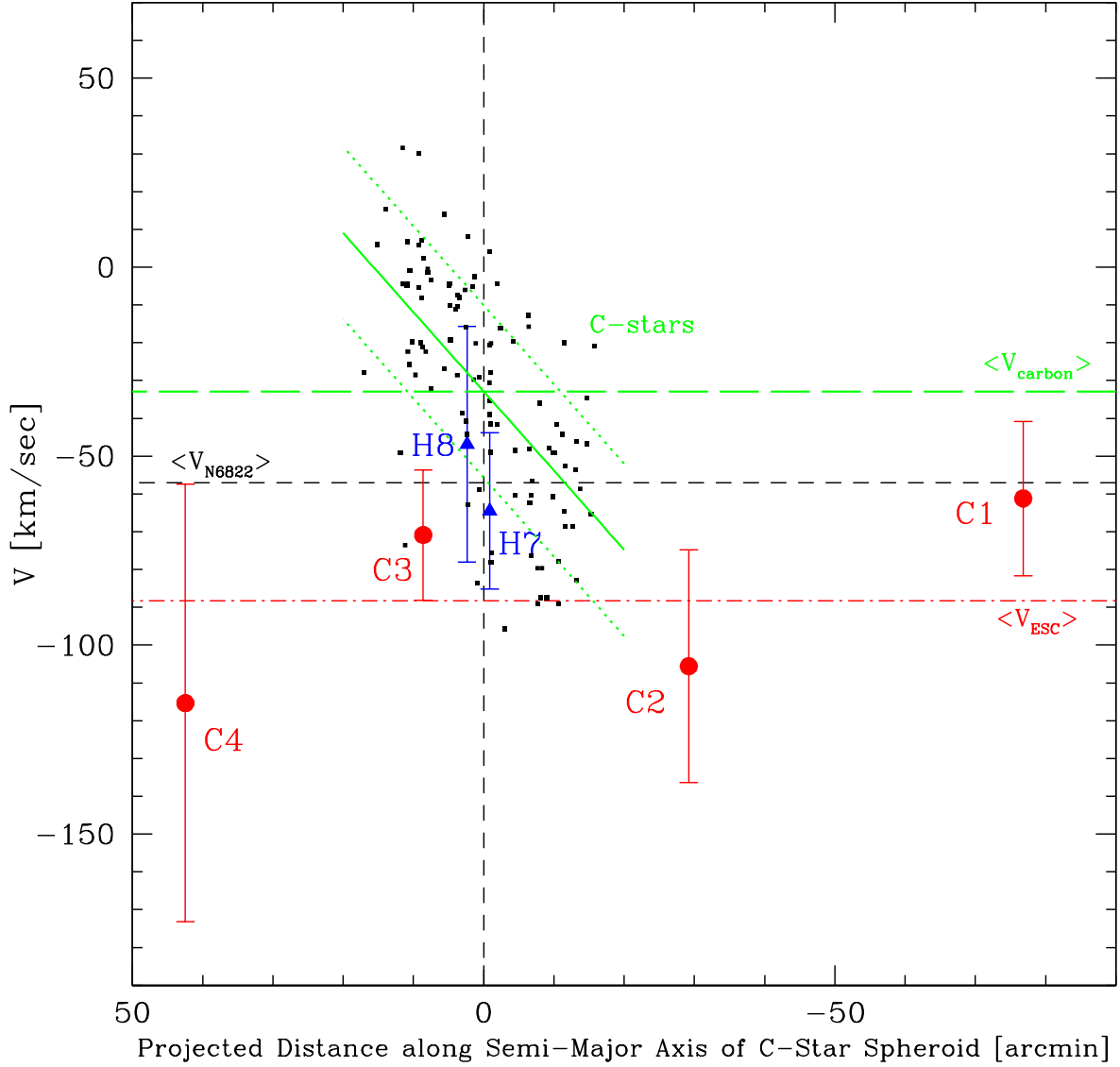}
 \caption{Line of sight velocity distribution of four ESCs as well as Hubble VII (H7) and Hubble VIII (H8) versus projected distances along the major axis of the carbon star spheroid of NGC 6822 defined in \citet{dem06}. The systemic velocity of NGC 6822 is marked by a horizontal dashed line at $-57$ km s$^{-1}$. The long-dashed line indicates the mean radial velocity of carbon stars ($-32.9 \pm 22.8 $ km s$^{-1}$) rotating with $v_{rot}=2.1 \pm 0.3$ km s$^{-1}$ arcmin$^{-1}$ or $15.9 \pm 0.3$ km s$^{-1}$ kpc$^{-1}$, and the dot-dashed line indicates the mean radial velocities of ESCs ($-88.3 \pm 22.7$ km s$^{-1}$).}
\label{kin1}
\end{figure}

Figure \ref{kin1} displays the radial velocities of ESCs and Hubble clusters as well as those of carbon stars along the projected distance from the NGC 6822 center. It is known that those carbon stars exhibit a systematic rotation around the main body of NGC 6822 \citep{dem06}.
Figure \ref{kin1} shows clearly that ESCs in NGC 6822 do not fit into the scheme of rotation displayed by carbon stars in the halo of NGC 6822 \citep{dem06}.
It is also evident that the velocities of ESCs do not show any sign of systematic motion either, while the velocities of two Hubble clusters may be consistent with the rotation of carbon stars: the velocities of H VII and H VIII are within $1.5 \sigma$ of the mean velocity of carbon stars. 
This result suggests that the ESCs are not kinematically coupled with the intermediate stellar populations and that the ESC system are pressure supported, rather than rotation supported.  

We have determined the mean velocity and velocity dispersion of carbon stars and ESCs:  $<v>= -32.9$ and $v_{\sigma} = 29.4$ km s$^{-1}$ for carbon stars, and  $<v> = -88.3$ and $v_{\sigma} = 22.7$ km s$^{-1}$ for ESCs. However, the carbon stars are systematically rotating so that we have derived the rotation velocity of carbon stars as $v_{rot} = 2.1 \pm 0.3$ km s$^{-1}$ arcmin$^{-1}$ or $15.9 \pm 0.3$ km s$^-{1}$ kpc$^-{1}$ based on the radial velocity data provided by \citet{dem06}. Then, the rotation corrected velocity dispersion of carbon stars is $v_{\sigma} = 22.8$ km s$^{-1}$.
Considering that the systematic radial velocity of NGC 6822 is $-57.0 \pm 2.0$ km/s as listed in NED,
the system of ESCs is moving toward us faster than NGC 6822 itself and carbon stars, while the velocity dispersion of ESC system is very similar to that of carbon stars.

These results show that ESC system has kinematical properties distinct from those of the other stellar components in NGC 6822 including carbon stars.
The HI-disk like structure does not have any common kinematic characteristic with ESC system in NGC 6822, either \citep{wel03}.
Considering the absence of any systematic motion, it may be a clue supporting that ESCs had accreted into the halo of NGC 6822, which is separated from the galaxy-wide kinematics involving the carbon stars and the HI-disk.

\subsection{Age and Metallicity of NGC 6822 ESCs}
\label{chem}

\begin{figure}
\plotone{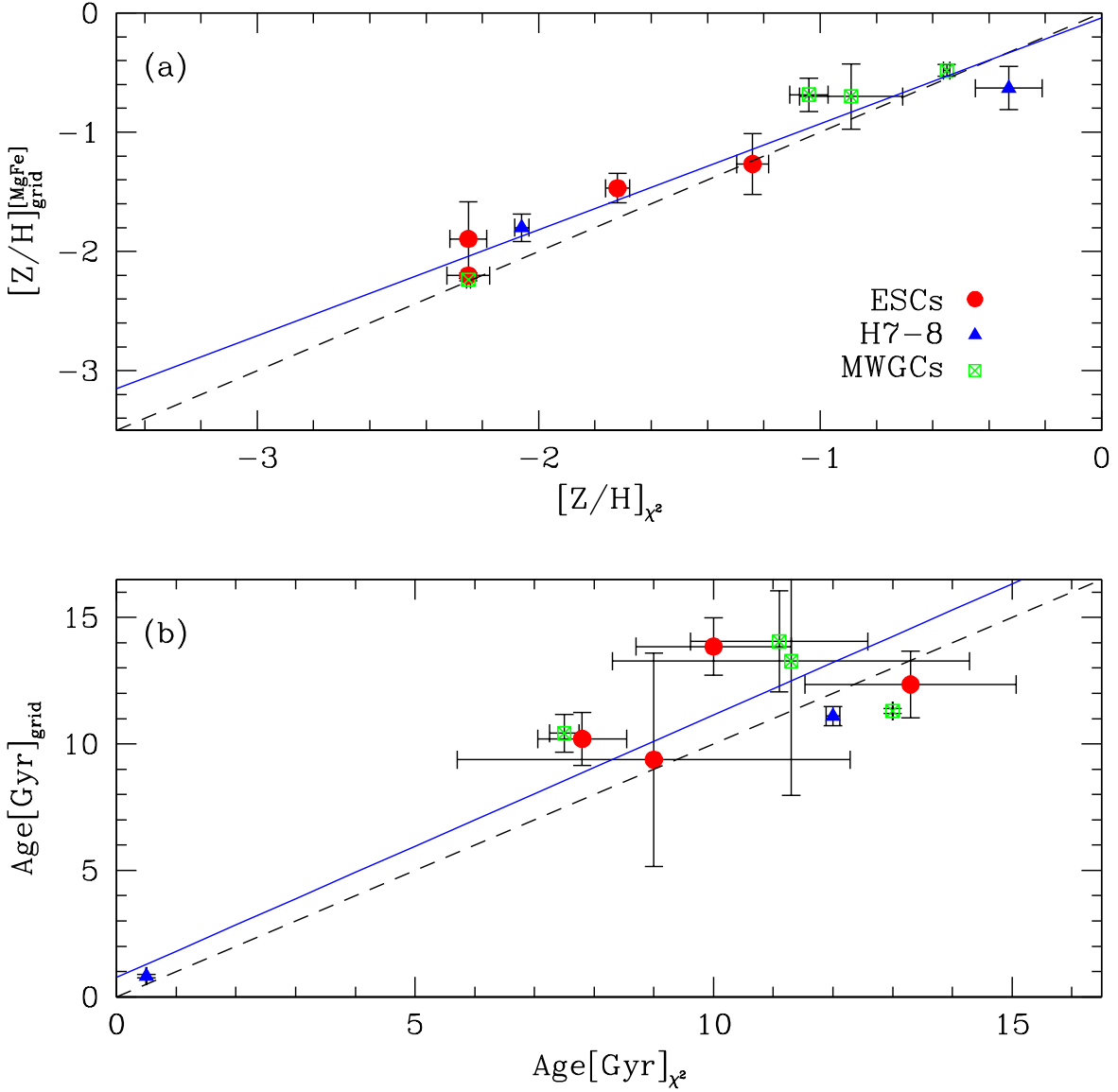}
\caption{Comparison of metallicities (a) and ages (b) derived by the grid method and the $\chi^2$ fit method. NGC 6822 ESCs are marked in circles, Hubble clusters are in triangles, and Milky Way globular clusters are in squares. Solid line indicates the result of linear least square fit, while dashed line marks the one-to-one correspondence. See text for details.}
\label{grid2chi}
\end{figure}

We list the measured metallicities and ages of each cluster in NGC 6822 in Table \ref{tab4}, and compare those values derived from the grid and $\chi^2$ fit methods in Figure \ref{grid2chi}.
The upper panel shows that both methods return consistent results on metallicity [Z/H]. We have derived linear least square fits between two metallicity values and found that the rms values are not greater than $0.2$. The lower panel also shows that the ages determined by the grid and $\chi^2$ methods are very well correlated with each other although there are some scatters with large errors. The result of linear least square fit exhibits an offset between two age measurements of about $1.4$ Gyr, which is usually smaller than the age errors marked in the plot.

\begin{figure}
\plotone{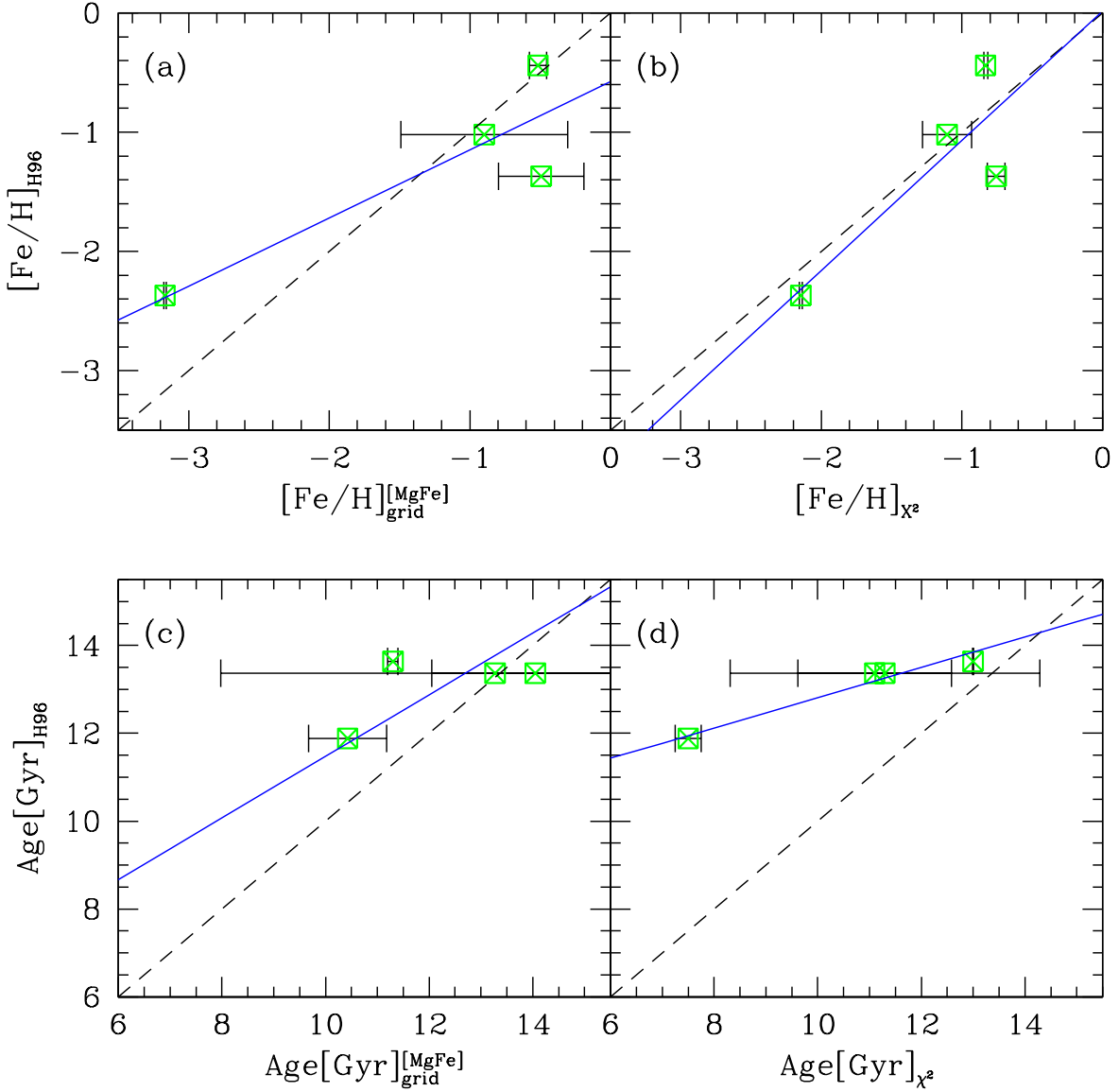}
\caption{Comparison of metallicities (panels a and b) and ages (panels c and d) for four MW GCs (NGC 6624, M12, M15, M107) with the values listed in the catalog by \citet{har96}. Solid line indicates the result of linear least square fit, while dashed line marks the one-to-one correspondence. See text for details. }
\label{mwgc}
\end{figure}

\begin{figure}
 \plotone{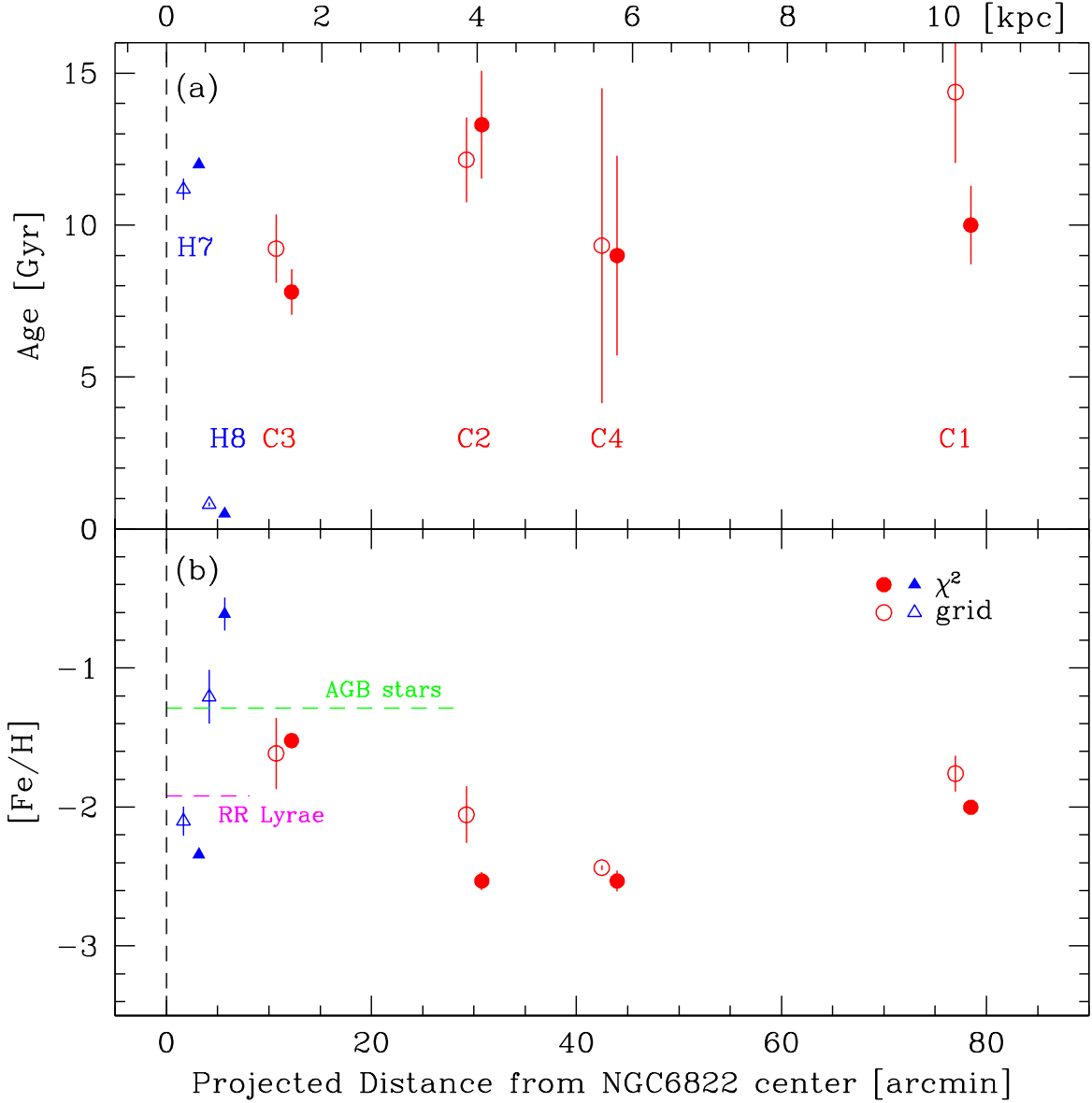}
 \caption{The radial distribution of age (panel a) and metallicity (panel b) for NGC 6822 star clusters. ESCs (C1-C4) are marked in circles, and H VII and H VIII are marked in triangles. Open symbols indicate the results derived by the grid method and filled symbols represent the values determined by the $\chi^2$ fit method. The mean metallicities for RR Lyrae \citep{cle03} and AGB stars \citep{sib12} as well as their sampled radial coverages are also marked in the lower panel by dashed lines.}
\label{agemet1}
\end{figure}

Figure \ref{mwgc} shows the comparison of age and metallicities of MW GCs derived by the grid and $\chi^2$ fit methods with those values listed in \citet{har96} (2010 version). We have assumed [$\alpha$/Fe]=0.3, the average value for MWGCs, to calculate the metallicities with the equation [Fe/H] $= [Z/H] - 0.94 \times {\rm [\alpha/Fe]}$ \citep{tho03}. It is noted that ages are relatively better reproduced by the grid method, while metallicities are better traced by the $\chi^2$ fit method. This result of comparison also suggests that the derived age of $7 - 8$ Gyr may actually be an indication of 10 Gyr or older due to the uncertainties larger than expected by the grid and $\chi^2$ fit methods themselves. Therefore, we adopt the age and metallicity values derived by the $\chi^2$ fit method for further discussions unless specifically noted otherwise, while we interpret the age with a caution. However, we keep ages and metallicities derived from both methods in Table \ref{tab4} and in Figure \ref{agemet1} for reference.

Figure \ref{agemet1}(a) displays the age distribution of 4 ESCs and 2 Hubble clusters in NGC 6822, determined by the $\chi^2$ fit (filled symbols) and the grid (open symbols) method. The result shows that all star clusters are older than about 8 Gyr except for H VIII that is estimated to be $<1$ Gyr old. It is noted that the age of H VII is estimated to be about 12 Gyr, while C2 is about 13 Gyr and C1 is estimated about 10 Gyr old. Considering that the oldest age that can be measured through the current theoretical model is about 10 -- 14 Gyr, and that H VII is known as a genuine old GC in NGC 6822 through the existing spectroscopic study \citep{coh98}, this result supports that ESCs C1 and C2 are almost as old as H VII, and C1 is a genuine old GC with the largest separation of about 77 arcmin or 10.5 kpc from the center of NGC 6822.

C3 is about 7.8 Gyr old although C3 is located relatively closer to NGC 6822 center than the other ESCs and has higher chance of younger stellar populations from the galaxy body. This implies that C3 may be slightly younger than H VII but still old, especially considering the uncertainty in the age determination. The age of C4 is about 9.0 Gyr with rather large error of about 3 Gyr. This large error may be due to the extended structure with low surface brightness of C4, resulting in low S/N spectra. Combined with its integrated color $(V-I)_0 = 1.12$ from \citet{hwa11}, C4 is probably old with $>$ 8 -- 10 Gyr in spite of the large age error.

It is noted that the age of H VIII is estimated younger than the other clusters of NGC 6822 investigated in this study: $0.5 \pm 0.01$ Gyr with the $\chi^2$ fit and $0.81 \pm 0.07$ Gyr with the grid method. The age measurements made by individual H lines return $0.46^{+0.10}_{-0.06}$ Gyr for H$\beta$, $1.02^{+0.10}_{-0.60}$ Gyr for H$\delta$, and $0.88^{+0.40}_{-0.15}$ Gyr for H$\gamma$ lines. Therefore, H VIII may be slightly older than indicated by the grid and the $\chi^2$ fit methods, maybe up to about $1.28$ Gyr. However, it still suggests that H VIII is younger than the other GCs in NGC 6822 including H VII and four ESCs. This result on the age of H VIII is in qualitative agreement with the result by \citet{cha00} that derived $1.4^{+1.1}_{-0.6}$ Gyr based on the optical spectroscopy made with CTIO and another report by \citet{wyd00} that derived $1.8 \pm 0.2$ Gyr based on the optical photometry data obtained with HST. On the other hand, \citet{str03} reports that the age of H VIII is about 3--4 Gyr based on the low resolution spectroscopy made with Keck, which is much older than the other results mentioned above.

Figure \ref{agemet1}(b) displays the radial metallicity distribution of 4 ESCs and 2 Hubble clusters in NGC 6822, calculated by assuming [$\alpha$/Fe]=0.3, the average value for MWGCs. This shows that ESCs and H VII are very metal poor with [Fe/H]$\approx -1.5 \sim -2.5$, ranging from [Fe/H]$= -2.53 \pm 0.08$ for C4 to [Fe/H]$= -1.52 \pm 0.06$ for C3. For H VIII, the metallicity is estimated to be [Fe/H]$= -0.61 \pm 0.12$. This metallicity for H VIII is higher than [Fe/H]$= -1.58 \pm 0.28$ derived for H VIII by \citet{str03}. 
However, the metallicity for H VII, [Fe/H]$=-2.34 \pm 0.03$ is consistent with the values in the literature, such as [Fe/H]$= -1.95 \pm 0.15$ by \citet{coh98} and [Fe/H]$=-1.79 \pm 0.25$ by \citet{cha00}.
The metallicity of C1 is [Fe/H]$= -2.00 \pm 0.04$, which suggests that the outermost ESC of NGC 6822 is a genuinely metal poor population.

The result on the age and metallicity in this study of NGC 6822 star clusters shows that star clusters older than about 8 Gyr are metal poor with [Fe/H]$\lesssim -1.5$, and that the old and metal poor clusters are preferentially located in the outer halo, farther than 15 arcmin or 2 kpc from the center of NGC 6822 except for H VII. Combined with the radial distribution of carbon stars shown in Figure \ref{kin1}, this suggests that the old stellar halo of NGC 6822, located roughly more than 20 arcmin or 2.6 kpc away from the galaxy center, is composed of old ($>$ 8 Gyr) and metal poor ([Fe/H]$\lesssim -2.0$) stellar populations. Compared with this, the inner halo of NGC 6822 is populated with intermediate age ($\sim 1 - 3$ Gyr) and slightly metal rich ([Fe/H]$\approx -0.6$) as well as old ($\sim 12$ Gyr) and metal poor ([Fe/H]$\approx -2.0$) populations.

\section{Discussion}

\subsection{ESCs as Tracers of Old Stellar Structure}

The existence of an old stellar spheroidal or halo around a dwarf galaxy and its physical properties have been subjects of many studies. For NGC 6822, an extended structure of old and intermediate age stars has been first detected by \citet{let02}. Further investigation by \citet{lee05} and \citet{dem06} revealed that the spheroidal or halo structure is extended out to, at least, about 30 arcmin from the center of NGC 6822.

The mean metallicity of stars in NGC 6822 is [Fe/H]=$-1.0 \pm 0.3$ based on the near infrared photometry of RGB stars \citep{dav03}, which is in good agreement with [Fe/H]=$-1.0 \pm 0.5$ determined by the equivalent widths of CaII triplet lines \citep{tol01}. However,  it is noted by \citet{tol01} that the metallicity ranges from -0.5 to -2.0 for stars selected in the central 5 to 11 arcmin region of NGC 6822. This metallicity spread is also consistent with [Fe/H]$=-1.92 \pm 0.35$ for old RR Lyrae variables \citep{cle03} and [Fe/H]$= -0.49 \pm 0.22$ for A-type supergiants \citep{ven01} in the central region of NGC 6822. Recently, \citet{sib12} show that the global metallicity of stars is [Fe/H]$=-1.27 \pm 0.07$ within about 28 arcmin radius from NGC 6822 center, hinting an existence of systematically lower metallicity component in the outer halo.

The star clusters of NGC 6822 investigated in this study are old ($>8-10$ Gyr) and metal poor ([Fe/H]$\lesssim -1.5$) except for one cluster: H VIII is relatively young ($< 1$ Gyr) and metal rich ([Fe/H]$>-1.0$). 
It is noted that the metallicities of very old clusters, C1, C2, and H VII, range from $-2.0$ to $-2.5$, which is in agreement with [Fe/H]$=-1.92 \pm 0.35$ for old RR Lyrae variables \citep{cle03}. This makes ESCs a good tracer of old stellar population.
Considering the wide spatial distribution of ESCs, this result suggests the presence of old and metal poor stellar halo extended out to as far as 77 arcmin or 10.5 kpc (e.g., C1) from the center of NGC 6822, one of the largest old stellar halos known for dwarf galaxies in the Local Group. This extent goes beyond the previously explored area \citep{let02,dem06} and covers about three times larger radial distance than observed by \citet{sib12}.

It is interesting to find out that, for NGC 6822, the extended old and metal poor population appears to exhibit different kinematics when compared with the intermediate age components. Figure \ref{kin1} shows that the radial velocities of ESCs do not fit into the rotation by carbon stars \citep{dem06}. Considering that carbon stars with about 1 -- 3 Gyr old are concentrated near the galaxy center, the central component or main body of NGC 6822 may be kinematically detached from the extended stellar halo mostly represented by ESCs. It is also noted that the rotation direction of Carbon stars are roughly consistent (approaching west wing and receding east wing) with the rotation curve of HI gas overlaid on the main body but extended toward the perpendicular direction to the stellar halo of NGC 6822 \citep{let02,wel03,dem06}. The distinct kinematic characteristics of different components in NGC 6822 may be the evidence of merging-like events \citep{deb00}, combined with different spatial distributions of different populations \citep{komi03,lee05}.

\subsection{Origin of ESCs}

Recent discoveries of many ESCs (e.g., \citealt{hwa11,hux11}) have led to inquiries regarding the origin of these ESCs and how they are made larger than typical GCs.
There have been discussions on the possible origins of these ESCs and some of the proposed scenarios are
summarized in \citet{hwa11}. Among them, there are scenarios involving tidally stripped dwarf galaxies, collisions of star clusters, and intrinsically extended star clusters that have survived tidal disruption.
It has been also implied that observed ESCs may be a mixture of heterogeneous populations
with different formation mechanisms.

The spectra of ESCs in NGC 6822 confirms that they are old stellar population with low metallicities. 
The large separation of these ESCs from the galaxy center also implies that 
they may be survivors of weak tidal force in the outer halo of NGC 6822.
These observational results are consistent with a theoretical model by \citet{hur10} that ESCs are natural 
products of early star formation in small dwarf galaxies with moderate to weak tidal forces, which is an ideal 
condition for ESCs to survive the tidal disruption. 
The relative isolation of NGC 6822 itself could have helped the formation and survival of ESCs in their current locations.

Even in the MW, it has been noted that large and extended GCs are preferentially found in the outer halo 
where the distance from the MW center, $R_{GC}$ is larger than 40 kpc \citep{vdb04}. Many of those 
extended GCs belong to the young halo (YH) population, which leads to a suggestion that these extended 
GCs may have come from the accreted dwarf galaxies \citep{zin93,mac04}. 
Considering the similar physical properties of ESCs in NGC 6822 to those extended GCs in MW, this may be 
independent evidence supporting the hypothesis that extended GCs mostly came from the accreted dwarf galaxies.
However, it is also argued that extended GCs in the outer part of MW-like galaxies can be explained by introducing the expansion of initially compact GCs, without invoking accretions of dwarf galaxies \citep{gie11,mad12} .

It is interesting to note that ESCs in NGC 6822 do not exhibit any clear signature of systematic kinematics 
unlike the intermediate age carbon stars, mostly concentrated around the central main body of the galaxy.
This indicates that ESCs, old population in the halo, are kinematically detached from the central body of NGC 6822, composed of young and intermediate age populations as well as a huge disk-like HI structure \citep{komi03,deb00}. 
Combined with the lack of any systematic kinematics among ESCs, this may suggest that ESCs in NGC 6822 have accreted from the external components such as neighboring dwarf galaxies.

The external origin of ESCs also implies that the observed ESCs may not be necessarily bound to NGC 6822 system. The best example can be the outermost ESC, C1. It should make C1 the nearest intergalactic globular cluster. 
At the moment, however, there is no clear evidence to confirm or to deny the dynamical containment of C1 or the other ESCs in NGC 6822 system. Even if we are to confirm that the ESCs are gravitationally bound to NGC 6822, it does not necessarily mean that they have originally formed in NGC 6822 or that they have accreted from other neighboring systems because the old ages and low metallicities are easily expected in both scenarios.

At least, it seems clear that ESCs are kinematically separated from intermediate stellar structure (e.g., carbon stars) as well as HI-disk like structure with young stars. Since old RGB stars exhibit a similar spatial distribution with ESCs, it may be interesting to investigate the kinematical and chemical properties of old RGB and RR Lyrae stars observed near the ESCs in NGC 6822 halo. With spectra of many such old stars, it will be easier to determine the overall kinematical properties of any halo stellar structures that might be associated with ESCs. The existence of such stellar structures with distinct kinematical properties, regardless of whether they are gravitationally bound to the host galaxy or not, could be construed as a result of the unrelaxed recent merger-like interactions of NGC 6822 with its neighbors.

Therefore, it will be very useful to compare the kinematical as well as chemical properties of ESCs with those of adjacent halo stars in many galaxies to solve the issue of the ESC origin as well as the dwarf galaxy accretion, even in the context of evolution of large MW-like galaxies.

\subsection{Dynamical Mass of NGC 6822}

We have estimated the dynamical mass of NGC 6822 by using the ``tracer mass estimator" method based on the assumption that the ESC system is pressure supported as a whole because no significant signature of rotation is observed.
We have adopted the definition of mass estimator defined by \citet{eva03} as  eq.\ref{eq1} combined with eq.\ref{eq2}, which is
described as a function of line of sight velocities $v_{los}$ and galactocentric distances $r$ of clusters:

\begin{equation}
M_{est} = \frac{C}{GN} \sum_{i=1}^{N} v^2_{los{\rm i}} r_{\rm i}
\label{eq1}
\end{equation}
where
\begin{equation}
C = {4(\alpha\!+\!\gamma) \over \pi}
    {4\!-\!\alpha\!-\!\gamma\over 3\!-\!\gamma}
    {1\!-\!(\rin/\rout)^{3-\gamma} \over
     1\!-\!(\rin/\rout)^{4\!-\!\alpha\!-\!\gamma}}.
\label{eq2}
\end{equation}

The $\gamma$ is determined by the tracer population density distribution and the $\alpha$ is an indicator of the properties of galaxy potential under consideration. When $\gamma=3$, which is for the typical stellar spheroidal halo, the value of C is given by
\begin{equation}
C = {\ds 4(\alpha\!+\!3)(1\!-\!\alpha) \over \ds \pi} 
{\ds \log (\rout/\rin)\over \ds 1\!-\!(\rin/\rout)^{1\!-\!\alpha}}.
\end{equation}

We also assume an isothermal potential for NGC 6822 halo, which makes $\alpha=0$. For comparison, two extreme cases with velocity distribution of tracers are tried: $C=16/\pi$ for an isotropic velocity distribution and $C=32/\pi$ for tracers on radial orbits and the estimated mass for each case is quoted as an upper and lower limit of the estimated mass hereafter.

\begin{figure}
 \plotone{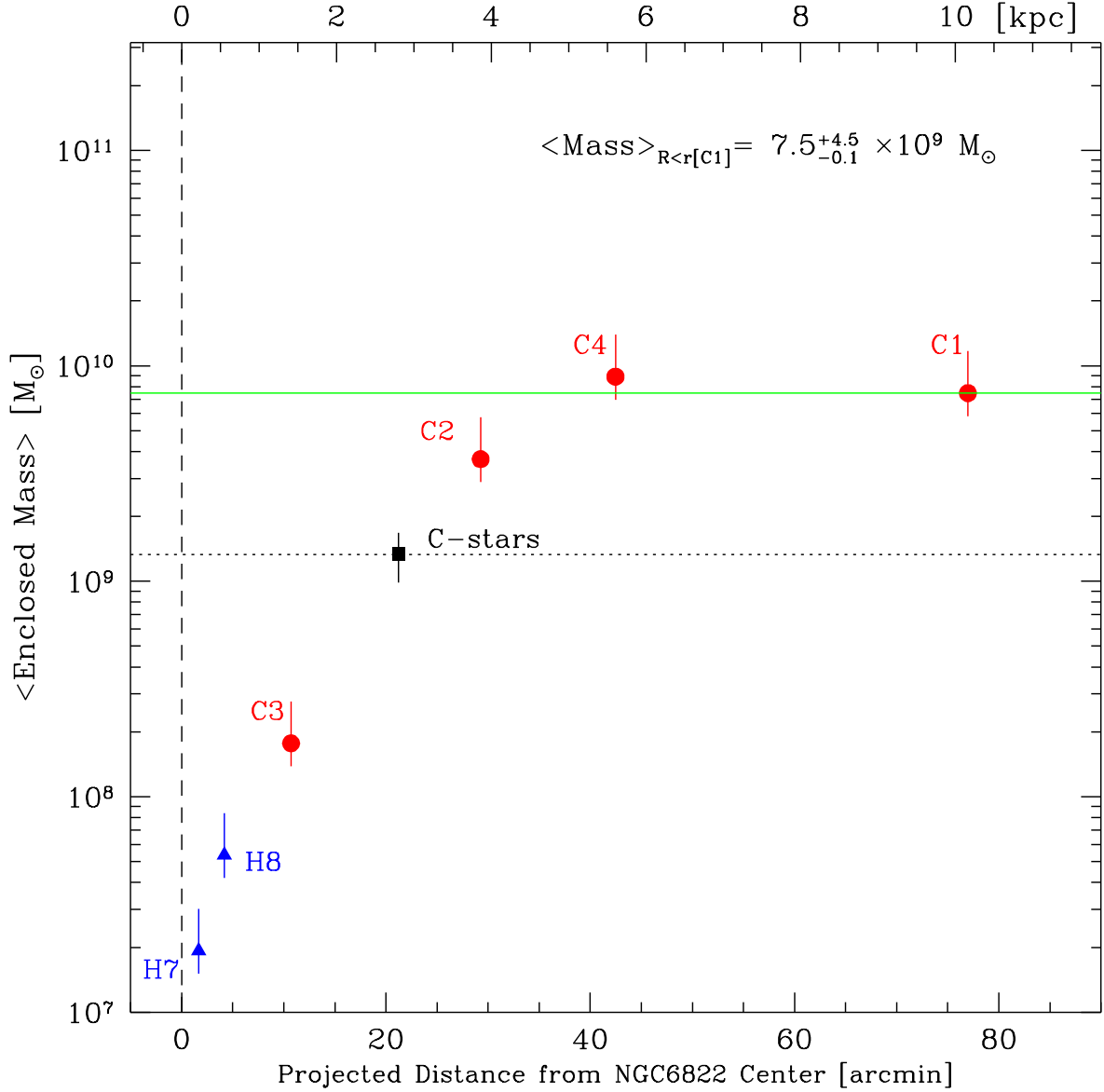}
 \caption{
 The average mass of NGC 6822 belonging to the inside of each star cluster. For Carbon stars, the mass is estimated at the outermost Carbon stars assuming a circular rotation (marked in a square). Note that the average mass of NGC 6822 system converges to $7.5^{+4.5}_{-0.1} \times 10^{9} M_{\odot}$ for $r>45$ arcmin. ESCs are marked in circles and Hubble clusters are in triangles.}
\label{kin2}
\end{figure}

Figure \ref{kin2} displays the estimated mass of NGC 6822 system calculated by assuming each cluster is the outermost tracer. The mass estimated by assuming a circular rotation of carbon stars is also marked for comparision.
The estimated mass increases from $1.8^{+1.0}_{-0.4} \times 10^8 M_{\odot}$ to $7.5^{+4.5}_{-0.1} \times 10^{9} M_{\odot}$ as we move from C3, the innermost ESC, to C1, the outermost ESC in NGC 6822. If we limit the radial range to include only out to C2, then the mass estimate is $3.7^{+2.1}_{-0.8} \times 10^9 M_{\odot}$. This value is still larger than the mass estimate made with carbon stars under the rotational velocity field, which is $1.3 \pm 0.4 \times 10^9 M_{\odot}$. However, it is noted that the mass estimate converges to about $7 - 9 \times 10^{9} M_{\odot}$ between C4 and C1, that is, between 40 and 77 arcmin from the center of NGC 6822.
We have determined the total dynamical mass of NGC 6822, M$_{N6822}=7.5^{+4.5}_{-0.1} \times 10^{9} M_{\odot}$, measured at the location of C1, about $77$ arcmin or 10.5 kpc from the galaxy center.
This mass is about twice the mass estimated from the HI rotation measurements combined by the pseudo-isohtermal (ISO) dark matter model assumption by \citet{wel03}, which is $3.2 \times 10^{9}~{\rm M_{\odot}}$ enclosed within about r$=30$ arcmin radius.

We have derived the M/L value of NGC 6822 using the newly determined dynamical mass and found that (M/L)$_{N6822} = 75^{+45}_{-1} (M/L)_{\odot}$. This M/L value is higher than previously known for NGC 6822: (M/L)$=17$ (M/L)$_{\odot}$\citep{mat98} and (M/L)$= 32$ (M/L)$_{\odot}$ \citep{wel03}. It is also higher than M/L of other dwarf galaxies in the Local Group with similar luminosities, i.e., $M_{V} \approx -15.0$, as shown in Figure 9 of \citet{mat98}: M/L is usually lower than 30 for dwarf galaxies with $M_{V} \approx -15.0$. It is noted that M/L of NGC 6822 is comparable with that of faint dwarf spheroidal galaxies such as Ursa Minor with $M_{V}=-8.9$ and (M/L)$=79 $ (M/L)$_{\odot}$ \citep{mat98}. This result indicates that NGC 6822 is one of the most dark matter dominated dwarf galaxies in the Local Group, which is also consistent with the conclusion made by \citet{wel03} based on the HI rotation curve.

\section{Summary and Conclusions}

The spectroscopic study of the four ESCs as well as two Hubble clusters in  NGC 6822 has shown that the radial velocities of ESCs range from about $-61.2$ km s$^{-1}$ (for C1) to about $-115.3$ km s$^{-1}$ (for C4), and the mean radial velocity of ESCs are $-88.3 \pm 22.7$ km s$^{-1}$, approaching toward us faster than NGC 6822 itself ($v_{los} = -57$ km s$^{-1}$) as well as system of carbon stars ($-32.9 \pm 22.8$ km s$^{-1}$). Interestingly, the ESCs do not display any sign of systematic rotation unlike carbon stars rotating within 20 arcmin from NGC 6822 center. The ages and metallicities derived using the Lick indices suggest that the ESCs are old ($\geq 8$ Gyr) and metal poor ([Fe/H]$\lesssim -1.5$). 
The metallicity of C1, the outermost ESC of NGC 6822,  is estimated to be [Fe/H]$= -2.00 \pm 0.04$, while the metallicity of C3, the innermost ESC, is [Fe/H]$=-1.52 \pm 0.06$. However, the lowest metallicity of [Fe/H]$\approx-2.5$ is expected from C2 and C4.

There is no clear sign of radial metallicity gradient in the star clusters of NGC 6822, but it is found that both metal poor ([Fe/H]$\approx -2.0$) and metal rich ([Fe/H]$\approx -0.9$) components co-exist within 15 arcmin or 2 kpc from the center of NGC 6822, while only metal poor component is observed in the outer halo with $r \geq 20$ arcmin or 2.6 kpc.
Considering no systematic motion of ESC system, we suggest that ESCs are primordial stellar components with low metallicity that had accreted into the outer halo of NGC 6822. The extended structure may have originated under the environment of relatively weak tidal force in a small dwarf galaxy, which is consistent with the theoretical predictions.

Based on the wide spatial distribution of ESCs, we have determined that the dynamical mass of NGC 6822 is M$_{N6822}=7.5^{+4.5}_{-0.1} \times 10^{9} M_{\odot}$ enclosed within 77 arcmin or 10.5 kpc in projected distance.
The M/L ratio of NGC 6822 is $75^{+45}_{-1} (M/L)_{\odot}$, which is much higher than previously known \citep{mat98} and about twice higher than estimated from the HI disk rotation curve \citep{wel03}.
This strongly indicates that NGC 6822 is a highly dark matter dominated system among the other dwarf galaxies in the Local Group. 

It is not clear yet whether the high content of dark matter has anything with formation and/or survival of ESCs in NGC 6822 halo. However, with more kinematical and chemical data for old stars in NGC 6822 halo, we may be able to better understand the origin of ESCs as well as the evolutionary history of NGC 6822 itself, a small dwarf galaxy with complex structure.

\acknowledgements

The authors are grateful to the anonymous referee for the useful
comments and suggestions that helped to improve the original manuscript.
This work was supported by the National Research Foundation of Korea (NRF) grant
funded by the Korea Government (MEST) (No. 2012R1A4A1028713).
N.H. was supported in part by Korea GMT Project operated by Korea Astronomy and Space Science Institute (KASI).
The authors are grateful to Thomas Puzia for generously providing the GONZO package and to Serge Demers for radial velocity data of NGC 6822 carbon stars.
This research has made use of the NASA/IPAC Extragalactic Database (NED) which is operated by the Jet Propulsion Laboratory, California Institute of Technology, under contract with the National Aeronautics and Space Administration.



\begin{deluxetable}{lc}
\tablecaption{Radial velocities of ESCs (C1--C4), and Hubble VII and VIII in NGC 6822.\label{tab1}}
\tablewidth{0pt}
\tablehead{ \colhead{ID} & \colhead{V$_{radial}$} \\
 & \colhead{km/sec} }
 \startdata
 NGC6822C1 & ~-61.2  $\pm$ 20.4
\\
 NGC6822C2 & -105.6 $\pm$ 30.7
\\
 NGC6822C3 & ~-70.9 $\pm$ 17.3
\\
 NGC6822C4 & -115.3 $\pm$ 57.9
\\
 Hubble VII & ~-64.5 $\pm$ 20.6
\\
 Hubble VIII & ~-46.9 $\pm$ 31.2
\\
 \enddata

\end{deluxetable}


\begin{deluxetable}{cccc}
\tablewidth{0pt}
\tablecaption{Zero point offsets for Lick index calibration.\label{tab2}}
\tablehead{ \colhead{index} & \colhead{unit} & \colhead{offset\tablenotemark{a}} & \colhead{rms} }
\startdata
   CN$_1$     & mag &  -0.084   & 0.056  \\
   CN$_2$     & mag &  -0.079   & 0.062  \\
   Ca4227     & \AA &  -0.140   & 0.989  \\
   G4300      & \AA &  -2.621   & 1.145  \\
  Fe4383      & \AA &   0.843   & 0.863  \\
  Ca4455      & \AA &  -0.283   & 1.002  \\
  Fe4531      & \AA &  -0.408   & 1.375  \\
  Fe4668      & \AA &  -1.886   & 1.932  \\
  H$\beta$    & \AA &   0.573   & 1.028  \\
  Fe5015      & \AA &  -0.657   & 1.662  \\
  Mg$_1$     & mag &  -0.001   & 0.055  \\
  Mg$_2$     & mag &   0.012   & 0.035  \\
  Mgb         & \AA &   0.091   & 1.071  \\
  Fe5270      & \AA &   0.125   & 1.051  \\
  Fe5335      & \AA &  -0.386   & 1.058  \\
  Fe5406      & \AA &  -0.171   & 1.006  \\
  Fe5709      & \AA &  -0.386   & 0.996  \\
  Fe5782      & \AA &   0.017   & 0.865  \\
  NaD      & \AA &   0.622   & 1.109  \\
  TiO$_1$     & mag &   0.000   & 0.035  \\
 H$\delta_A$      & \AA &   1.697   & 2.794  \\
 H$\gamma_A$      & \AA &  -1.338   & 0.937  \\
 H$\delta_F$      & \AA &   0.910   & 1.985  \\
 H$\gamma_F$      & \AA &   0.659   & 1.763  \\
\enddata
\tablenotetext{a}{Index(Lick)=Index(GMOS)+offset}
\end{deluxetable}


\begin{deluxetable}{lcccccccccccccccccccccccc}
\tabletypesize{\scriptsize}
\tabletypesize{\tiny}
\setlength{\tabcolsep}{0.035in}
\rotate
\tablecaption{Lick line indices of ESCs (C1--C4), and Hubble VII (H7) and VIII (H8) in NGC 6822.\label{tab3a}}
\tablewidth{0pt}
\tablehead{ \colhead{ID} & \colhead{CN$_1$} & \colhead{CN$_2$} & \colhead{Ca4227} & \colhead{G4300} & \colhead{Fe4383} & \colhead{Ca4455} & \colhead{Fe4531} & \colhead{Fe4668} & \colhead{H$\beta$} & \colhead{Fe5015} & \colhead{Mg$_1$} & \colhead{Mg$_2$} & \colhead{Mgb} & \colhead{Fe5270} & \colhead{Fe5335} & \colhead{Fe5406} & \colhead{Fe5709} & \colhead{Fe5782} & \colhead{NaD} & \colhead{TiO$_1$} & \colhead{H$\delta_A$} & \colhead{H$\gamma_A$} & \colhead{H$\delta_F$} & \colhead{H$\gamma_F$} \\
   & \colhead{mag}  & \colhead{mag} & \colhead{$\rm \AA$} & \colhead{$\rm \AA$} & \colhead{$\rm \AA$} & \colhead{$\rm \AA$} & \colhead{$\rm \AA$} & \colhead{$\rm \AA$} & \colhead{$\rm \AA$} & \colhead{$\rm \AA$} & \colhead{mag} & \colhead{mag} & \colhead{$\rm \AA$} & \colhead{$\rm \AA$} & \colhead{$\rm \AA$} & \colhead{$\rm \AA$} & \colhead{$\rm \AA$} & \colhead{$\rm \AA$} & \colhead{$\rm \AA$} & \colhead{mag} & \colhead{$\rm \AA$} & \colhead{$\rm \AA$} & \colhead{$\rm \AA$} & \colhead{$\rm \AA$} }
\startdata
C1 & -0.019 & -0.068 &  0.411 &  1.902 &  1.240 &  0.273 &  1.103 & -1.408 &  1.763 &  1.541 &  0.006 &  0.064 &  1.176 &  0.914 &  1.426 & -0.559 &  0.718 &  0.037 &  2.319 &  0.008 & 2.346 &  2.896 &  1.663 &  2.390  \\
C2 & -0.028 & -0.062 &  0.451 &  1.754 &  0.475 &  0.485 &  0.366 &  0.611 &  2.868 &  0.058 &  0.017 &  0.066 &  0.600 &  0.529 &  0.347 & -0.022 & -0.262 & -0.292 &  4.519 &  0.057 & 2.726 &  1.174 &  2.847 &  1.906  \\
C3 & -0.013 & -0.005 &  1.030 &  4.475 & -0.393 &  0.429 &  2.275 &  0.542 &  2.827 &  3.494 &  0.002 &  0.050 &  0.581 &  1.104 &  1.945 &  0.852 &  0.495 &  0.117 &  0.973 & 0.036 & -1.060 &  0.715 & -0.375 &  1.083  \\
C4 & -0.069 & -0.211 & -1.567 &  2.421 & -4.056 &  0.084 &  3.584 & -1.098 &  2.387 & -0.540 &  0.002 &  0.056 & -2.364 &  0.833 &  0.039 &  1.629 &  0.270 & -0.113 &  0.954 &  0.098 & 1.469 &  3.233 & -3.239 &  0.777  \\
H7 &  0.009 & -0.004 &  0.236 &  1.332 &  0.262 &  0.537 &  0.829 & -0.995 &  2.865 &  1.386 &  0.010 &  0.043 &  0.780 &  0.950 &  0.927 &  0.411 &  0.638 &  0.313 &  0.990 &  0.006 & 3.298 &  2.860 &  3.557 &  3.099  \\
H8 & -0.101 & -0.085 &  0.251 &  2.482 & -2.825 &  1.309 &  2.156 &  0.014 &  5.001 &  2.451 & -0.015 &  0.047 &  0.678 &  1.281 &  1.401 &  0.730 &  0.139 &  0.347 &  2.242 &  0.026 & 7.536 &  7.059 &  5.699 &  5.119  
\enddata
\end{deluxetable}

\begin{deluxetable}{lcccccccccccccccccccccccc}
\tabletypesize{\scriptsize}
\tabletypesize{\tiny}
\setlength{\tabcolsep}{0.035in}
\rotate
\tablecaption{Errors of Lick indices for ESCs (C1--C4), and Hubble VII (H7) and VIII (H8) in NGC 6822.\label{tab3b}}
\tablewidth{0pt}
\tablehead{ \colhead{ID} & \colhead{CN$_1$} & \colhead{CN$_2$} & \colhead{Ca4227} & \colhead{G4300} & \colhead{Fe4383} & \colhead{Ca4455} & \colhead{Fe4531} & \colhead{Fe4668} & \colhead{H$\beta$} & \colhead{Fe5015} & \colhead{Mg$_1$} & \colhead{Mg$_2$} & \colhead{Mgb} & \colhead{Fe5270} & \colhead{Fe5335} & \colhead{Fe5406} & \colhead{Fe5709} & \colhead{Fe5782} & \colhead{NaD} & \colhead{TiO$_1$} & \colhead{H$\delta_A$} & \colhead{H$\gamma_A$} & \colhead{H$\delta_F$} & \colhead{H$\gamma_F$} \\
   & \colhead{mag}  & \colhead{mag} & \colhead{$\rm \AA$} & \colhead{$\rm \AA$} & \colhead{$\rm \AA$} & \colhead{$\rm \AA$} & \colhead{$\rm \AA$} & \colhead{$\rm \AA$} & \colhead{$\rm \AA$} & \colhead{$\rm \AA$} & \colhead{mag} & \colhead{mag} & \colhead{$\rm \AA$} & \colhead{$\rm \AA$} & \colhead{$\rm \AA$} & \colhead{$\rm \AA$} & \colhead{$\rm \AA$} & \colhead{$\rm \AA$} & \colhead{$\rm \AA$} & \colhead{mag} & \colhead{$\rm \AA$} & \colhead{$\rm \AA$} & \colhead{$\rm \AA$} & \colhead{$\rm \AA$} }
\startdata
C1 &   0.002 &   0.004 &   0.143 &   0.168 &   0.192 &   0.195 &   0.203 &   0.224 &   0.226 &   0.238 &   0.007 &   0.007 &   0.244 &   0.246 &   0.248 &   0.249 &   0.249 &   0.250 &   0.251 &   0.007 &   0.280 &   0.297 &   0.303 &   0.307  \\
C2 &   0.005 &   0.008 &   0.269 &   0.289 &   0.328 &   0.333 &   0.357 &   0.396 &   0.400 &   0.413 &   0.012 &   0.012 &   0.424 &   0.426 &   0.429 &   0.430 &   0.431 &   0.432 &   0.433 &   0.012 &  0.479 &   0.490 &   0.500 &   0.505  \\
C3 &   0.004 &   0.006 &   0.221 &   0.240 &   0.274 &   0.280 &   0.294 &   0.318 &   0.320 &   0.329 &   0.010 &   0.010 &   0.343 &   0.345 &   0.347 &   0.348 &   0.349 &   0.350 &   0.351 &   0.011 &  0.387 &   0.401 &   0.415 &   0.420  \\
C4 &   0.013 &   0.020 &   0.720 &   0.833 &   0.951 &   0.990 &   1.035 &   1.118 &   1.126 &   1.159 &   0.035 &   0.035 &   1.191 &   1.197 &   1.206 &   1.211 &   1.214 &   1.216 &   1.221 &   0.037 &  1.383 &   1.452 &   1.503 &   1.523  \\
H7 &   0.001 &   0.002 &   0.073 &   0.079 &   0.092 &   0.093 &   0.097 &   0.109 &   0.109 &   0.112 &   0.003 &   0.003 &   0.115 &   0.116 &   0.116 &   0.117 &   0.117 &   0.117 &   0.118 &   0.003 &  0.130 &   0.133 &   0.138 &   0.139  \\
H8 &   0.005 &   0.007 &   0.249 &   0.268 &   0.322 &   0.327 &   0.343 &   0.373 &   0.375 &   0.389 &   0.010 &   0.010 &   0.403 &   0.406 &   0.410 &   0.411 &   0.413 &   0.413 &   0.415 &   0.011 &  0.459 &   0.474 &   0.484 &   0.489 
\enddata
\end{deluxetable}

\begin{deluxetable}{lcccccc}
\tablecaption{Ages and Metallicities of NGC 6822 ESCs (C1--C4), and Hubble VII and VIII\label{tab4}}
\tablewidth{0pt}
\tablehead{ \colhead{ID} & \colhead{Age$_{\rm grid}$} & \colhead{[Z/H]$^{\rm [MgFe]}_{\rm grid}$} & \colhead{[Fe/H]$_{\rm grid}$} & \colhead{Age$_{\rm \chi^2}$} & \colhead{[Z/H]$_{\rm \chi^2}$} & \colhead{[Fe/H]$_{\rm \chi^2}$} \\
 & \colhead{Gyr} & \colhead{dex} & \colhead{dex} & \colhead{Gyr} & \colhead{dex}  & \colhead{dex} }
\startdata
NGC6822C1 & $14.38 \pm  2.33$ & $-1.48 \pm  0.13$ & -1.76 & $10.00 \pm  1.30$ & $-1.72 \pm  0.04$ & -2.00 \\
NGC6822C2 & $12.15 \pm  1.40$ & $-1.77 \pm  0.20$ & -2.06 & $13.30 \pm  1.77$ & $-2.25 \pm  0.06$ & -2.53 \\
NGC6822C3 & $ 9.23 \pm  1.12$ & $-1.33 \pm  0.26$ &  -1.61 & ~$7.80 \pm  0.74$ & $-1.24 \pm  0.06$ & -1.52 \\
NGC6822C4 &  ~$9.33 \pm  5.18$ & $-2.15 \pm  0.02$ &  -2.44 & ~$9.00 \pm  3.29$ & $-2.25 \pm  0.08$ & -2.53 \\
Hubble VII & $11.18 \pm  0.35$ & $-1.82 \pm  0.11$ & -2.10 & $12.00 \pm  0.11$ & $-2.06 \pm  0.03$ & -2.34 \\
Hubble VIII &  ~$0.81 \pm  0.07$ & $-0.93 \pm  0.19$ & -1.21 & ~ $0.50 \pm  0.01$ & $-0.33 \pm  0.12$ & -0.61 \\
\enddata
\end{deluxetable}

\end{document}